

\hsize 174mm
\hoffset=-7mm
\vsize 235mm
\font\svt=cmr10 scaled \magstep3
\font\twelbf=cmb10 scaled \magstep1
\font\fif=cmr10 scaled \magstep2
\font\twelrm=cmr12
\font\twelit=cmti10 scaled \magstep1
{\svt
\hskip 135mm {\twelrm TOHO-FP-9243}
\vglue 2cm
\centerline{ Massive Analogue of Ashtekar-CJD Action}
\vglue 2cm}
{\fif
\centerline{ Kiyoshi Kamimura and Takeshi Fukuyama{$^\dagger$}}
\centerline{}
\baselineskip 20pt
\centerline{Department of Physics, Toho University}
\centerline{Funabashi, Japan 274}
\centerline{and}
\centerline{{$^\dagger$}Department of Physics, Ritsumeikan University}
\centerline{Kyoto, Japan 603}
\vglue 60mm}
\baselineskip 15pt
\twelrm
\centerline {abstract}
The action of Ashtekar gravity have been found by Cappovilla, Jacobson
and Dell. It does not depend on the metric nor the signature of the
space-time. The action has a similar structure as that of a massless
relativistic particle. The former is naturally generalized by adding a
term analogous to a mass term of the relativistic particle.
The new action possesses a constant parameter regarded as a kind of a
cosmological constant. It is interesting to find a covariant Einstein
equation from the action. In order to do it we will examine how the
geometrical quantities are determined from the non-metric action and
how the Einstein equation follows from it.
\vglue 2cm
----------------------------------------------------------------------

Presented at "Quantum Physics and the Universe"

(Waseda University, Tokyo, August 19-22, 1992)
\eject
\twelrm
\baselineskip 15pt
A new form of canonical gravity developed by Ashtekar~(1986,1987,1991)
has various nice features. It is formulated in a form of a gauge theory
with the gauge group SO(3). The constraints generating
the general coordinate and the gauge transformations are of polynomial
form and are suitable for canonical quantization.
However, the canonical variables are complex valued and must
satisfy the so called {\twelit reality conditions}. It is nicely
formulated starting from an action which is a sum of Einstein and
complex total divergence terms and the significance of reality
conditions has been discussed (Ashtekar (1986,1987),
Fukuyama and Kamimura (1990)).

The action in terms of SO(3) gauge fields is found by making
a Legendre transformation backward from the Hamiltonian to Lagrangian
formalism . It has been first given by Capovilla, Jacobson and
Dell~(CJD)~(1989). The CJD action for the pure gravity theory is
$${
{\cal L}_0=-{1\over{4 \eta}} [G^{ab}G_{ab}-{1\over 2}G^a_{\; a}G^b_{\; b}
],} \eqno(1) $$
where $G_{ab}$ is a building block of the action defined in
terms of $SO(3)$ gauge field strength,
$${
G_{ab}={1\over 4}\epsilon^{\mu\nu\rho\sigma}F_{\mu\nu a}
F_{\rho\sigma b},~~~~F_{\mu\nu a}=
\partial_{[\mu}A_{\nu]a}+g\epsilon_{abc}A_{\mu}^{\; b} A_{\nu}^{\; c}.}
\eqno(2) $$
$\eta$  in the Lagrangian is a scalar density multiplier field.
The action is invariant under general coordinate transformations as well
as $SO(3)$ gauge transformations. A characteristic property of the
action is that it does not depend on any space-time metric. Especially
there is {\twelit no a priori} sign of signature of the metric
(Euclidian or Lorentzian). $\epsilon^{\mu \nu \rho \sigma} $ is simply
a Levi-Civita symbol, $\epsilon^{0123}=1$. Another is absence of the
reality conditions which are imposed in the Ashtekar formalism.

In the previous paper (Kamimura, Makita and Fukuyama, 1992)
we have discussed how the space-time metric arises from the algebraic
properties of the constraints and the Hamiltonian.
It is accomplished by finding a set of first class constraints satisfying
the same algebra as diffeomorphism constriants.
The metric of equal time space appears in the structure function.
The temporal components of the metric are found in the
coefficients of the diffeomorphism constraints in the Hamiltonian.
There is one ambiguity in determining the weight factor of deformation
generator of normal direction.
The signature of the metric is determined by the choice of the reality
conditions. Possible forms of reality conditions on the canonical
variables are determined by a requirement that the derived tetrad and
spin connections are real quantities. The $SO(3)$~ gauge connection is
shown to be a self-dual spin connection both in Euclidian and Lorentzian
signatures. It is complex valued in the Lorentzian case.
In the quantum theory the signature factor may be taken into a choice
of the inner product (Fukuyama and Kamimura 1990).

The CJD action is generalized to those including cosmological constants
(Bengtsson (1990) Perdan (1991)).They discussed a Hamiltonian constraint
$${
  H_\Lambda = {1\over 2} \epsilon_{abc} \epsilon_{ijk}(\pi^{ia}
\pi^{jb} B^{kc}       -{\Lambda} \pi^{ia}\pi^{jb}\pi^{kc})=0,}
\eqno(3) $$
which is obtained by adding a cosmological constant term to the Ashtekar
constraint for the pure Einstein theory. Since the constraint is third
order polynomial with respect to the momenta the inverse Legendre
transformation results in a Lagrangian of non simple form.
It does not seem to be a {\twelit fundamental Lagrangian of the gravity}.

An additional constant is introduced into the CJD action
by referring to the case of a relativistic particle.
The action of a relativistic free particle is given by
$~~{ L_2=-\mu{\sqrt{{\dot x}^2}},}~~$
where $\mu$ is a constant having the meaning of mass of the particle.
It is rewritten using a scalar density Lagrange multiplier~ $\eta$~ as
$~{ L_1={-{1 \over{4 \eta}}}{\dot x}^2-{\eta }\mu^2.}~~$
In this form we can take the smooth massless limit and obtain an
action for the massless particle
$~{ L_0={-{1 \over{4 \eta}}}{\dot x}^2.}~~$

 The CJD action (1)~ has a similar structure to the massless particle
action. It is naturally generalized by adding a term, which is
analogous to the mass term in the case of a relativistic particle, as
$${
{\cal L}_1={-{1\over {4 \eta}}} [G^{ab}G_{ab}-{1\over 2}G^a_{\; a}
G^b_{\; b}]-\eta {\mu}^2 .}
\eqno(4) $$
Here the constant $\mu$ is dimensionless for the conventional assignments
of the gauge fields. For non vanishing value of $\mu$~ we have an
action of square root form through the elimination of $\eta$
$${
{\cal L}_2={-\mu}\sqrt{~G^{ab}G_{ab}-{1\over 2}G^a_{\; a}G^b_{\; b}~}.}
\eqno(5) $$
The actions in ~(4)~ and ~(5)~ may be classically equivalent though the
branch of square root in the latter must be examined carefully.
In this way we can introduce an additional constant without violating
any local symmetries of the original action.
The action (4)~ has a smooth $\mu \rightarrow 0$~ limit to (1)~ so
it is a {\twelit neighbour} of the pure Einstein theory (Bengtsson 1990).
In order to examine roles of the constant $\mu$~ and find out how the
Einstein equation is modified by it, we must find out the
geometrical quantities as functions of the canonical variables.
In this paper we present an explicit derivation of Einstein Equation
from the CJD action (1)~ in a form applicable to general case
including the action of present interest (4).

In the Hamiltonian formalism of the action ~(4)~ there appears a set of
first class constraints,
$${\pi_\eta = 0, }~~~{\pi^{0a} = 0, }
\eqno(6)$$
$$
{J^a = D_i \pi^{ia} = 0,}
\eqno(7) $$
$$
{T_j = \pi^{ka} F_{jka}=0}
\eqno(8) $$
and
$${
H_0 = {1\over 2} \epsilon_{abc} \epsilon_{ijk}(\pi^{ia} \pi^{jb} B^{kc}
                        +{{2\mu^2} \over 3} B^{ia} B^{jb} B^{kc})=0,}
\eqno(9) $$
where $\pi_\eta$ and $\pi^{\mu a}$ are momenta conjugate to $\eta$ and
$A_{\mu a}$ respectively, and $D_\mu$ is $SO(3)$ covariant derivative.
The magnetic field is defined by $B^{ia} \equiv {1\over 2}\epsilon^{ijk}
F_{jk}^{\quad a}$. The constraints in (6) tell that
{}~$\eta$~and~$A_{0a}$~ are arbitrary multipliers. Equation (7)~ is the
Gauss law constraint for $SO(3)$. The last two of them,
(8) and (9),~are reflecting the general covariance of the Lagrangian.
The $\mu^2$ dependent term appears only in the Hamiltonian constraint
{}~$H_0$. It has a smooth limit to the Ashtekar constraint.

The same set of constraints (7-9) are obtained starting from the action
(5). In this case the Hamiltonian constraint $~H_0=0~$ arises as the
primary constraint rather than the secondary one. It is more natural
that both the Hamiltonian and the momentum constraints appear in the
same stage of the canonical formalism.

The Hamiltonian constraint (9)~ is compared with the one in (3).
The $~det \pi~$ term in (3)~ is replaced by $~det B~$ in (9).
At first glance they seem to describe quite different systems.
This point must be carefully examined since identification of the
dynamical variables to the geometrical variables is not same for these
cases. In the case of (3), the momentum is identified with the
densitized triad and the gauge field is regarded as the self-dual spin
connection as in the zero cosmological constant case. On the other
hand the correspondence is modified in the case of (9).

The canonical Hamiltonian  ${\cal H} = p \dot q - {\cal L}$~  becomes
$${
{\cal H}=\pi^{0a}\dot A_{0a} -A_{0a}J^a+ {1\over 2}\epsilon^{ijk}E_{ia}
  {\bf B}_j^{\; a}T_k+{\eta\over{2detB}}H_0,}
\eqno(10) $$
where the electric field is $E_{ia} \equiv F_{0ia}$ and
${\bf B}_{ia}$ is the inverse of $B^{ia}$ and $det B (\equiv det B^{ia})$
is assumed to be non vanishing. In~(10)~ we have used only
{}~ $ ( p - {{\partial L} \over{\partial \dot q}}  )^2 = 0$~as strong
equality but ~ $ ( p - {{\partial L}\over{\partial \dot q}}  )= 0$~has
never been used. By this prescription we can obtain correct forms of
multipliers on constraints in terms of $p,q$ and $\dot q$ (Kamimura,1982).

The space-time metric is introduced if we can identify
the constraints with diffeomorphism generators $H_\perp$ and $H_j
(j=1,2,3)$. In a metric space $H_j$'s generate
transformations of three coordinates of a space-like hyper-surface.
$H_\perp$, on the other hand, deforms the hyper-surface in its normal
direction $n_\mu$ . They satisfy the following algebra (Teitelboim 1973).
$$
  {\lbrace H_i(x),\; H_j(y)\rbrace =H_i(y){\partial\over \partial x^j}
  \delta (x-y)-H_j(x){\partial \over \partial y^i}\delta(x-y), }
\eqno(11)$$
$$
  {\lbrace H_j (x),\; H_\perp (y)\rbrace =H_\perp (x){\partial
  \over \partial x^i}\delta(x-y) }
\eqno(12)$$
and
$$
  {\lbrace H_\perp (x),\; H_\perp (y) \rbrace =-\epsilon(\gamma ^{ij}(x)
  H_j(x)+\gamma^{ij}(y)H_j(y)){\partial \over \partial x^i}\delta(x-y),}
\eqno(13)$$
where $\epsilon = n_\mu n^\mu$ is a signature of the metric {\twelit
i.e.}~ $\epsilon= +1 $ ~for Euclidian signature and $-1$ for Lorentzian
one. $\gamma^{ij}$~ is the induced metric of the space-like hyper-surface
and appears in the Poisson bracket (13) as the structure function.

The constraints of the present system satisfy the same form of algebra
$$
  {\lbrace T_i(x),\; T_j(y) \rbrace =T_i(y){\partial \over \partial x^j}
  \delta(x-y)-T_j(x){\partial \over \partial y^i}\delta(x-y)},
\eqno(14)$$
$${
  \lbrace T_j(x),\; {1\over h^{1/2}}H_0(y)\rbrace
  ={1\over h^{1/2}}H_0(x) {\partial \over \partial x^j}\delta (x-y)}
\eqno(15)$$
and
$${
  \lbrace {1\over h^{1/2}} H_0(x) , {1\over h^{1/2}} H_0(y) \rbrace =
  -[{{\pi^{ia}\pi^j_{\>a}-2\mu^2 B^{ia}B^j_{\>a}} \over h} T_j(x) +
    {{\pi^{ia}\pi^j_{\>a}-2\mu^2 B^{ia}B^j_{\>a}} \over h} T_j(y)
    ]{\partial \over \partial x^i}\delta(x-y).}
\eqno(16)$$
\vskip 3mm
Here and thereafter equalities are satisfied up to the $SO(3)$ gauge
constraint; $J^a=0$. It means that additional $SO(3)$ transformations
are associated in the commutators. The weight factor $h$ ~multiplied
on $H_0$~ is a first order homogeneous function of $det \pi,~ det B$~
and $\epsilon_{ijk}\epsilon_{abc}\pi^{ia}B^{jb}B^{kc}$~ and
is assumed to be non vanishing. The case of  $h=0$~ corresponds
to a degenerate metric, which is not  prohibited in the system of action.
When $h$ vanishes, in a certain region of parameter space of $x^\mu$~,
the metric (of weight zero) is not well defined there.
The action may describe such a generalization of the Einstein theory .

By comparing (11-13)~ with (14-16)~  we find
$$
H_j~=T_j,~~~~~  {H_\perp = {1\over h^{1/2}}H_0,}~~~~~
{\gamma^{ij}=\epsilon~ {{\pi^{ia}\pi^j_{\; a}-2\mu^2 B^{ia}B^j_{\;a}}
\over h}.}\eqno(17)$$
The identification of $H_\perp$~ and $\gamma^{ij}$~ is not unique since
$h~$ has not been determined.

The remaining components of the metric are found by
noting that the Hamiltonian is the generator of surface deformation along
the time axis . By projecting a vector along the time axis into normal
and tangential components of the equal time surface the lapse and shift
functions are determined as
$$
 { \alpha \equiv{1\over{\sqrt{\epsilon g^{00}}}} =
  {{\epsilon \eta  h^{1\over2}}\over{2 \; detB}}}, ~~~~
  {{\beta}^k \equiv -{g^{0k}\over g^{00}}
  ={1\over2} \epsilon^{ijk} E_{ia}{\bf B}_j^{\;a}}
\eqno(18)$$
and $$
  {\gamma^{ij}=g^{ij}-{{g^{0i}g^{0j}}\over{g^{00}}}.}
\eqno(19)$$

We have written all the components of the metric in terms of the
dynamical variables of the action, the signature factor $\epsilon$
and the density weight factor $h$. We will examine whether the metric
$g^{\mu\nu}$ determined from (17-19) acquires a consistent
interpretation. The requirements for the metric are  the following:
{}~~1) all the components are real quantities and ~2) $\epsilon g^{00}$
must be positive. ~3) $\gamma^{ij}$ is positive definite so that the
equal-time space spans a space-like hyper-surface.

The lapse function $\alpha$ in (18)~is a multiplier of {\twelit
secondary} first class constraint $H_\perp$~ and is taken to be
positive as a result of the gauge fixing condition on $\eta$. The
shift vector ${\beta}^j$ ~is a multiplier of {\twelit primary} first
class constraint $T_j$~and is taken to be real as the gauge choice.
$\gamma^{ij}$, on the other hand, is determined dynamically by (17)~.

In the following we will show how the Einstein equation comes out
explicitly. First we introduce an (inverse) triad variable $e^{ia}~$
by taking a square root of $\gamma^{ij}~$.
Next we introduce a spin connection  $\omega_{\mu AB}(A=0,1,2,3)$~ as
an auxiliary function. It is defined using a torsion free condition
$${
 \partial_{[\mu}e_{\nu]A} + \omega_{[\mu A}^{\;\quad B} e_{\nu ] B}=0}
\eqno(20)$$
with the tetrad variables $e_{\mu A}$,
$${
e_{ia}={(e^{ia})^{-1}},
\quad e_{00}=\alpha,\quad
e_{0a}=\beta^je_{ja},\quad e_{i0}=0.}
\eqno(21)$$
{}From the
torsion free condition the components of the spin connection are solved
as functions of $\pi^{ia},\; B^{ia},\; \alpha,\; \beta^j,\;
{\dot\pi}^{ia}$~ and $\dot B^{ia}$~.
Among those, $\omega_{iab}$ is given as a  function of $\pi^{ia}~$ and
$B^{ia}$ only,
$${
 \omega_{iab}=-{1\over 2}e_i^{\;c}(A_{abc}-A_{bca}-A_{cab});
 \qquad A_{abc} \equiv e^j_{\; a}e^k_{\; b} \partial_{[j}e_{k]c}.}
$$
Other components, $\omega_{i0a},\; \omega_{0ab}$~ and $\omega_{00a}$,~
depend on the time derivative of $\pi^{ia}$~and $B^{ia}~$ . They are
evaluated using Hamilton's equation of motion with the Hamiltonian (10)~
$${
 {\cal H}=\pi^{0a}\dot A_{0a} -A_{0a}J^a+ {\beta^j}T_k+
         \epsilon \alpha H_{\perp}.}
\eqno(21)$$
The Hamilton's equations are
$$
 {\dot\pi^{ia}=D_j[{{\epsilon \alpha} \over {h^{1\over2}}}\epsilon^{abc}
 (\pi^j_{\;b}\pi^i_{\;c}+2\mu^2 B^j_{\;b}B^i_{\;c})]+D_j[\beta^j\pi^{ia}
-\beta^i\pi^{ja}]+g\epsilon^{abc}\pi^i_{\;b}A_{0c},}
\eqno(22)$$
$$
\dot A_{ia}={{\epsilon \alpha} \over {h^{1\over2}}}
            \epsilon_{abc}\epsilon_{ijk}\pi^{jb}B^{kc}
           -\epsilon_{ijk}\beta^j B^k_{\;a}+D_i A_{0a}.
\eqno(23)$$
Thus the components of the spin connection are given as functions of
{}~$\pi^{ia},~ \alpha,~ \beta^j$~ and ~$A_{\mu a}$~(and their spatial
derivatives).

In the case of $\mu=0$~, the first equation of motion (22) means that
the gauge connection is the self dual spin connection. In showing it,
it is crucial that the density weight factor $h$~ is chosen to satisfy
$$
D_i{{e_{ia}}\over{det~e_{jb}}}=0.
\eqno(24)$$
It is satisfied from the Gauss law constraint (7)
if $h$~ is proportional to $det\pi$ .
The proportionality constant is chosen so that the tetrad becomes
a real quantity,
$$
h^{1\over2}={1\over I}({\kappa det\pi})^{1\over 2}~~\longrightarrow~~
e^{ia}={\pi^{ia}\over{(\kappa det\pi)}^{1\over 2}}.
\eqno(25)$$
where  $I=1$~ for the Euclidian and $I=i$~ for the Lorentzian metric.
By using (22) the components of the spin connection  are explicitly
evaluated and expressed as
$${
  {Ig \over 2}A_{\mu a}={1\over 2}(\omega_{\mu 0a}-{I\over 2}\epsilon
  _{abc}\omega_\mu^{\;bc})\; \equiv \omega_{\mu 0a}^{(+)}}.
\eqno(26)$$
The self dual Riemannian tensor is then related to the gauge field
strength by
$$
  {Ig \over 2}F_{ij a}=R_{ij 0a}^{(+)}.
\eqno(27)$$
Using this the constraints (8) and (9) are written as
$$
\epsilon^{ijk}\epsilon^{abc}R_{ij 0b}^{(+)}e_{kc}=0,~~~~~~~~~
\epsilon^{ijk}R_{ij 0a}^{(+)}e_k^{\;a}=0.
\eqno(28)$$
The second equation of motion (23) is expressed in terms of self dual
Riemannian tensors as
$$
R_{0i 0a}^{(+)}+{I\alpha}\epsilon_{abc}e^{jb}R_{ij 0}^{~~~c(+)}
           -\beta^j R_{ij 0a}^{(+)}=0.
\eqno(29)$$
The equations (28) and (29)~ are exactly the set of Einstein equations
for pure gravity theory written in terms of self dual Riemannian  tensor.
Thus in the case of $\mu=0$ the pure Einstein gravity is reproduced from
the action (1). The density weight $h$~ has been determined under
the criteria (24) and the equivalence of this system with that given by
Ashtekar is completed.

For non-vanishing value of $\mu~$, we have not found a covariant form of
the Einstein equation. For a small value of $\mu^2$~ we can
treat terms of $\mu^2~$ perturbatively. For example the triad becomes
$$
e^{ia}={1\over{ h^{1\over 2}}}(\pi^{ia}-\mu^2 B^{ib}\pi_{jb}B^{ja}),
$$
where $\pi_{ia}~$ is the inverse of $\pi^{ia}$. Equations (24-29) will
be modified by terms of order $\mu^2$~. The preliminary result shows that
the additional terms do not mean presence of a standard cosmological
constant but suggest a higher order curvature and/or torsion theory.

\vskip 5mm

\centerline {REFERENCES}
\tenrm
\baselineskip 12pt
\settabs 1 \columns
\+Ashtekar A,(1986) Phys. Rev. Lett. {\twelbf 57}, 2244 ,
(1987) Phys. Rev. {\bf D36}, 1587,(1991) "Lectures on \cr
\+Non-Perturbative Canonical Gravity",(World Scientific, 1991).\cr
\+Bengtsson I, (1991) "Ashtekar's variables and the Cosmological
Constants", G\"oteborg preprint ITP91-33,(1991).\cr
\+Capovilla R, Jacobson T and Dell J, (1989) Phys. Rev. Lett.
{\twelbf 63} 2325.\cr
\+Fukuyama T and Kamimura K, (1990) Phys. Rev. {\twelbf D41}, 1105 ,
                            Phys. Rev. {\twelbf D41}, 1885 . \cr
\+Kamimura K,(1982) Nuovo Cimento {\twelbf 68B}, 33 .\cr
\+Kamimura K, Makita S and Fukuyama T, (1992) "Metric from non metric
action of gravity"\cr
\+~~~~preprint TOHO-FP-9242, to appear in Int. J. Mod. Phys. D.\cr
\+Perdan P,(1991) Class. Quant. Grav. 8, 1765. \cr
\+Teitelboim C, (1973) Ann. of Phys. {\twelbf 79}, 542 .\cr
\end